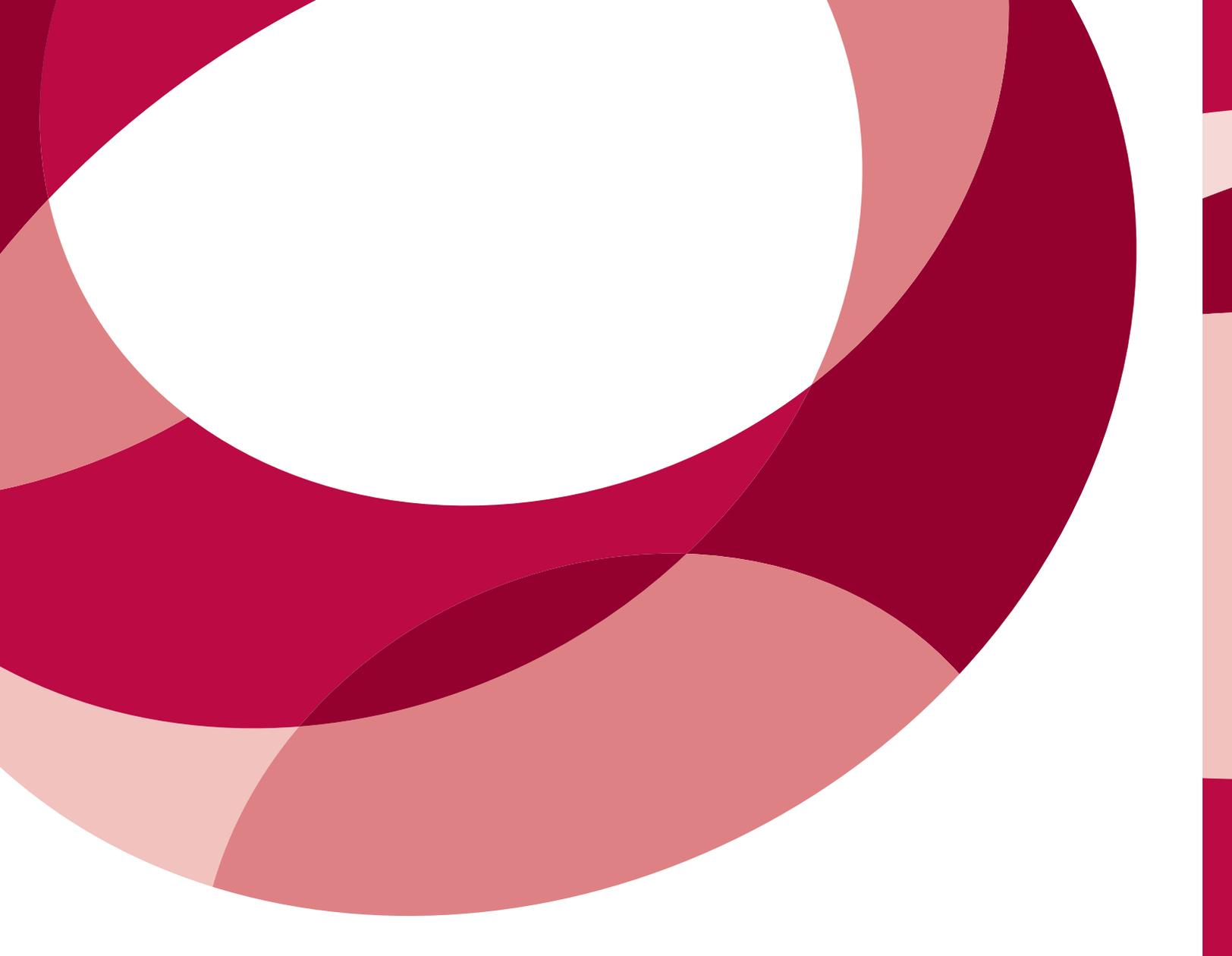

# Research Opportunities in Sociotechnical Interventions for Health Disparity Reduction

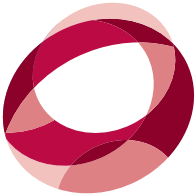

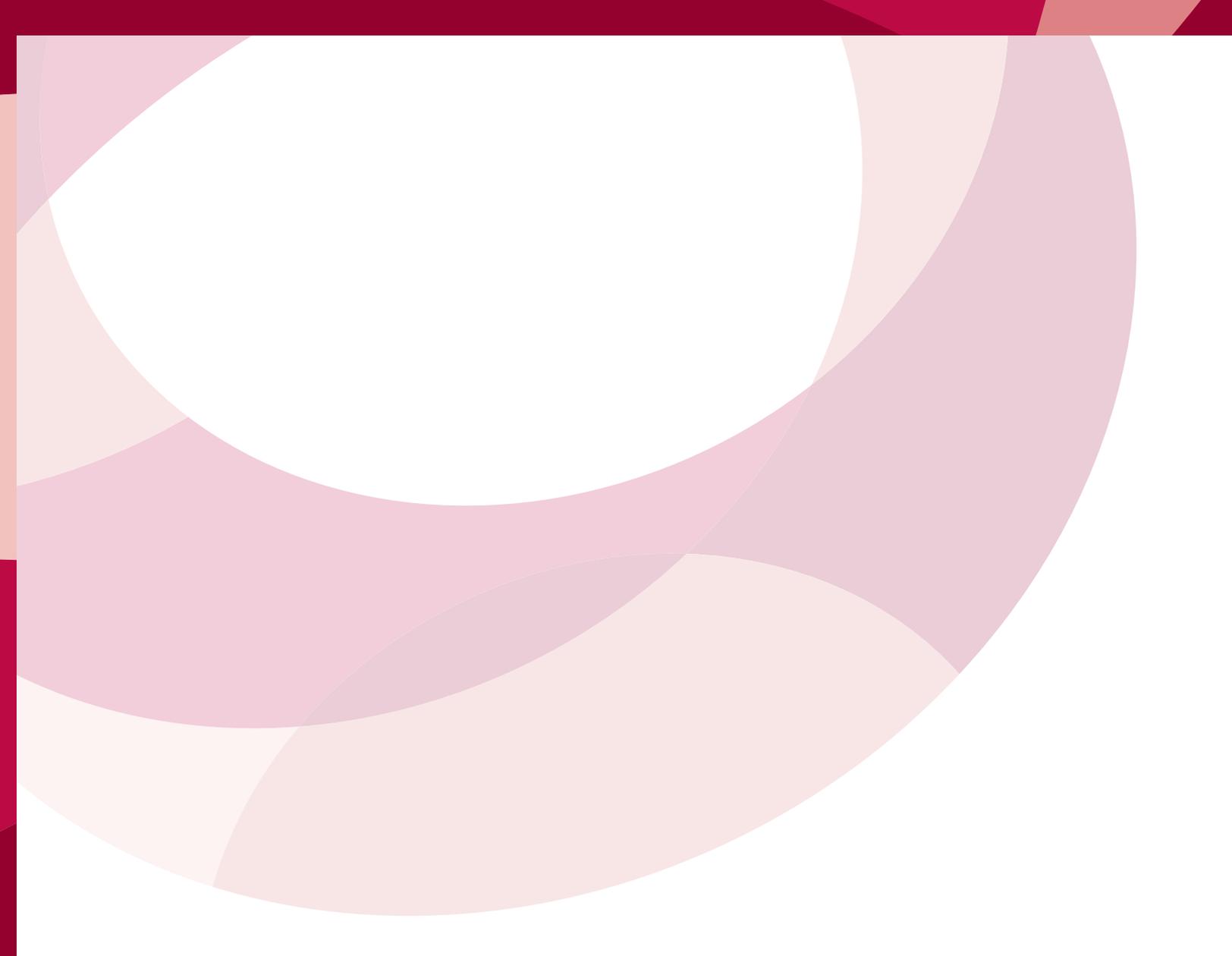

The authors gratefully acknowledge the contributions of the workshop planning committee members to the design and facilitation of the workshop, which formed the basis of this document. Committee members included Heather Cole-Lewis, Syed Haider, Eric Hekler, Predrag Klasnja and Donna Spruit-Metz.

This material is based upon work supported by the National Science Foundation under Grant No. 1734706. Any opinions, findings, and conclusions or recommendations expressed in this material are those of the authors and do not necessarily reflect the views of the National Science Foundation.

# Research Opportunities in Sociotechnical Interventions for Health Disparity Reduction

Katie Siek, Tiffany Veinot, and Beth Mynatt

June 2019

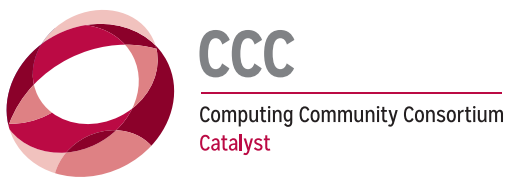

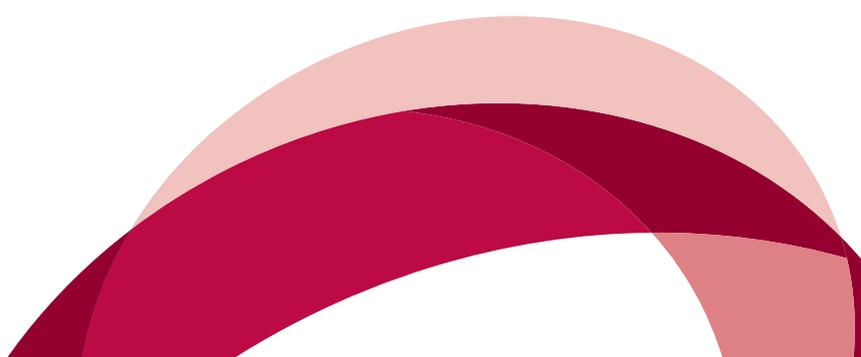

# 1. Overview

The implicit and explicit biases built into our computing systems [1] are becoming increasingly clear — they impact everything from targeting of advertisements [2] to how we are identified as people [3]. These biases disproportionately affect marginalized groups — people who are excluded from mainstream social, economic, cultural, or political life [4] — more acutely. While these biases can affect all aspects of our lives, from leisure [5] to criminal justice [6] to personal finances [7], they are all the more critical in the context of health and healthcare due to their significant personal and societal implications. In this interdisciplinary workshop, we explored how to design and build health systems for diverse populations through the following disciplinary lenses.

**Human computer interaction (HCI) researchers** address the growing need to empower lay populations to manage their health by designing, developing, and deploying novel sociotechnical interventions [8, 9]. As research in interactive systems in healthcare has matured, computing and health informatics researchers have increasingly drawn upon social and behavioral science theories [10] to design, develop, and analyze sociotechnical systems.

**Health informatics researchers** focus on basic research concerning patient information needs [11] and healthcare-oriented topics such as implementation of technologies in healthcare contexts, technical standards, health policy, impacts on healthcare quality, and access to, and uptake of, technologies [12]. Researchers also concentrate on the development of analytical techniques and algorithms focused on applied clinical problems such as illness diagnosis and prognosis.

**Behavioral medicine researchers** explore psychosocial mechanisms underlying health behavior — from determinants of behavior to how behavior is changed. Additionally, behavioral medicine research has a longstanding research focus on health disparities. At the same time, behavioral medicine researchers have traditionally developed health behavior theories and models through participant self-report or by utilizing commodity systems to evaluate the theory at scale.

**Health disparity researchers** investigate the prevalence and underlying correlates of health disparities, typically using observational study methods originating in epidemiology, such as cohort and case-control designs. Additionally, clinical epidemiologists contribute methods in the areas of research synthesis, with a recent focus on equity-focused systematic reviews that can inform intervention design [13-15].

Critically, reduction of health disparities (see box 1) through socio-technical interventions requires the knowledge and methods of each of these fields. Because health disparities are rooted in a variety of social, behavioral, economic, and healthcare-based factors, there is a need for researchers to consider the insights and research methods offered by each of these fields when designing and deploying interventions [16]. Furthermore, designing interventions that will be engaging to, and usable by, health disparity populations is a prerequisite for intervention impact — critical insights about which can be provided from different perspectives in each of these fields. Moreover, because interventions that could work well for health disparity populations may not be available to or readily adopted by them, there is a need to consider policy and implementation issues such as integration with healthcare systems and workflows, technology platforms, and incentives [16, 17] — challenges which researchers from these four fields are best positioned to tackle. There is also a need to incorporate understanding of the mechanisms driving different health disparities into design, implementation, and evaluation. Assessing the equity impact of interventions [18] in the context of specific studies, or across studies, is also critical.

The Computing Community Consortium (CCC) sponsored a two-day workshop titled Sociotechnical Interventions for Health Disparity Reduction in collaboration with the leadership of the Society for Behavioral Medicine's (SBM) 39th Annual Meeting on Monday, April 9 and Tuesday, April 10, 2018 in New Orleans, Louisiana. The workshop's goal was to bring together leading researchers in computing, health informatics, behavioral medicine, and health disparities to develop an integrative research agenda focused on sociotechnical interventions to reduce health disparities and improve the health of marginalized populations. The workshop was informed by four themes:

> **Health disparities are differences in disease prevalence, incidence, morbidity and/or mortality in one group as compared to the general population. In Western countries, groups which experience disparities in health outcomes include:**
> - People of lower socio-economic status (SES) based on income, wealth, education, and occupation;
> - Racial and ethnic minority groups including African Americans, Latinos, Native Hawaiians/Pacific Islanders, and Indigenous peoples;
> - Rural and urban residents;
> - Lesbian, gay, bisexual and transgender (LGBT) people;
> - People with disabilities; and
> - Men or women (varies by health issue).

Box 1: The definition of health disparities.

- **Theory to Design and Implementation:** Sociotechnical interventions that reduce health disparities require interdisciplinary knowledge to inform intervention design and implementation because health disparities are rooted in social, behavioral, economic, and healthcare-based factors.

- **Sociotechnical System Blackboxes:** As research in interactive systems in healthcare has matured, computing and health informatics researchers have increasingly drawn upon social and behavioral science theories to design, develop, and analyze sociotechnical systems. However, we do not always know why or how sociotechnical interventions "work."

- **Sociotechnical Systems to Inform Theory:** Behavioral medicine researchers have traditionally developed health behavior theories and models through participant self-report or by utilizing commodity systems to evaluate the theory at scale. However, data collected by sociotechnical systems can be leveraged more consistently to help develop existing behavioral science theories or extend new theories.

- **Multidimensional Evaluation to Reduce Health Disparities at the Population Level:** Sociotechnical interventions hold promise for reducing disparities and improving the health of marginalized populations, however interventions can generate unintended consequences that exacerbate disparities. There is a need to proactively evaluate equity impacts of sociotechnical interventions, at all phases of design and implementation.

The four themes were explored through two short multi-disciplinary panels and coordinated discussions followed by summarizing presentations to ensure that researchers from different disciplines had the opportunity to listen, learn, and share with each other. The researchers identified major research challenges and opportunities within each theme, specifically the need to:

- **Develop and evaluate equity-centered intervention strategies and implementation approaches.** Prevailing intervention strategies, which often focus on individual patient effort, behavior and choice may be less effective for marginalized populations — supporting a greater focus on upstream and multi-level interventions. Furthermore, existing approaches for implementing systems (e.g., promoting uptake and ongoing usage) tend to favor advantaged groups. There is a need for new approaches that can ensure equitable outcomes, as well as uptake and usage, of effective interventions.

- **Enhance participatory methods for designing, studying, and evaluating technology.** To ensure that we effectively address real problems, marginalized groups should be involved in choosing intervention priorities and designing and evaluating interventions. While researchers currently use participatory methods, there is a need to evaluate and improve these methods. Critically, there is a need to develop and support mechanisms for building capacity for marginalized communities to meaningfully participate in health research on socio-technical systems.

- **Build dynamic and multilevel theories for designing interventions.** Existing sociobehavioral theories typically

do not account for the dynamism of new types of sociotechnical systems. Moreover, few available theories have been developed with marginalized populations in mind, including the social and economic conditions that contribute to marginalization. Development of new theories can be facilitated by better mapping of theories onto sociotechnical systems and with new methodologies that can learn both new "hypotheses" and construct new theories or extend current theories using data.

- **Advance methods for dosing, tailoring, and optimizing sociotechnical interventions.** Little is known about how much usage of sociotechnical systems is needed to gain benefits from them and what doses of other aspects of the intervention context are needed to benefit from interventions (e.g., neighborhood walkability). Particularly for marginalized groups, we also know little about what aspects of interventions should change based on individual or contextual characteristics. Knowledge in this area can be gained through greater support for pilot studies, developing methods to strengthen conclusions based on small-N studies, and studies which tailor to characteristics relevant to health disparities.

- **Evaluate systems via multiple dimensions to reduce health disparities at the population level.** It is important for any sociotechnical intervention to be evaluated in relation to its impacts on health equity. Interventions should also be assessed at multiple levels where applicable (micro, meso, macro). Researchers should ask themselves equity-related questions (see Box 2 on page 18) in relation to any intervention studies and plan studies in which differential uptake, engagement and outcomes can be assessed. It is also important for researchers to examine potential unintended consequences — particularly through qualitative research. There is also a need for research and tools to assist researchers in evaluating the ethical implications of studies that gather data from marginalized participants, especially those that use third-party platforms and that capture social and community contexts.

- **Create interdisciplinary bridges to continue collaborating.** There is a need for development of a consortium or national centers to address health disparities with sociotechnical systems that creates a collaborative network of researchers, industry, providers, payers, and communities to aid in scaling sociotechnical interventions. This consortium should create reusable components, share algorithms and data, and develop approaches for transferability and robust partnerships.

This multidisciplinary workshop sought to take stock of prior successes and failures, of accumulated learnings and persistent challenges. In addition, workshop attendees sought to identify knowledge gaps and opportunities for advancement through research.

## 2. Developing Equity-centered Intervention Strategies and Implementation Approaches

For sociotechnical interventions to reduce health disparities, it is critical that intervention strategies — activities or features that aim to improve some predetermined health-related outcome — are grounded in an understanding of health disparities and the ways in which inequity can emerge at all stages of the intervention cycle, from access to effectiveness. This means that interventionists must understand what populations experience disparities for a given health outcome, the antecedents of those disparities, and potential theoretical pathways by which those disparities can be reduced. Workshop participants specifically advocated for the further development of "upstream interventions," described in section 2.1, to achieve this. In addition, interventions can only have an effect if they are adopted and used; or, in a research context, that participants are recruited and then remain in a study. Because marginalized groups are less likely to do these things, there is a need for equity-centered implementation approaches focused on adoption/recruitment and usage/retention.

### 2.1 Upstream Interventions

The extension of the World Health Organization's model on health disparities, shown in Figure 1, newly incorporates technology, meso-level factors found in other models, and disparities on the basis of LGBT identity, disability, and place of residence. As the left side of the model shows, the ultimate sources of health disparities can be found in macro-level factors associated with the sociopolitical and economic context. The model depicts four mechanisms of action by which social factors produce health disparities: (1) social stratification/marginalization (e.g., economic resources, power, prestige, residential and educational segregation, stigmatization and discrimination); (2) differences in exposures (e.g., environmental hazards, bullying); (3) differences in vulnerability; and (4) differences in social, economic, and health consequences once a person has become ill. Sociotechnical interventions focused on health disparity reduction can target any of these mechanisms, but those that move from the right to the left can be considered further "upstream."

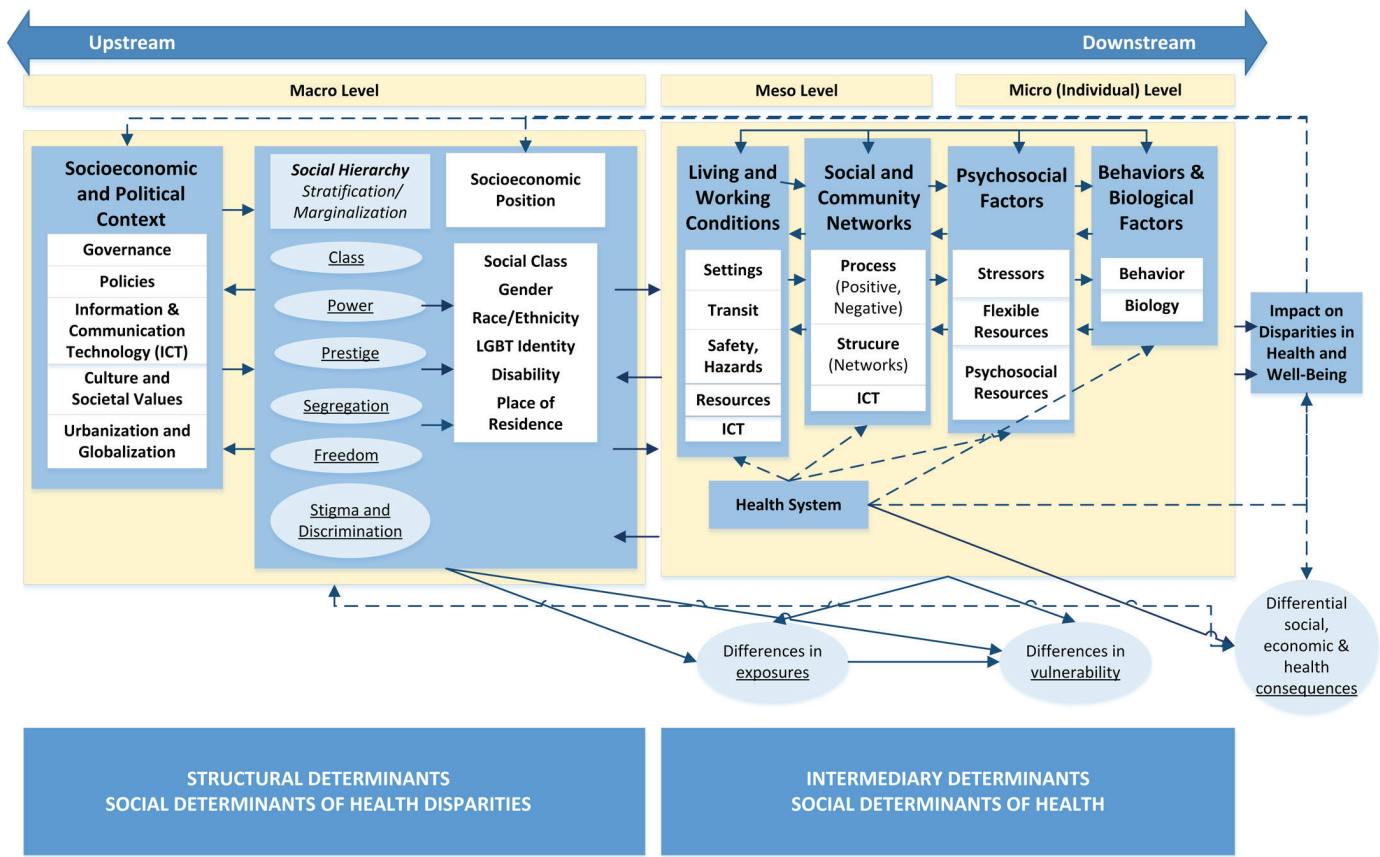

*Figure 1. Extension of the World Health Organization's model on Health Disparities [77]*

A limitation with current sociotechnical interventions is their overwhelming focus on downstream interventions, with most interventions designed to help individuals adopt healthy behaviors or better manage illness. However, equity-focused systematic reviews have shown that downstream health behavior interventions tend to be less effective for marginalized populations than those that operate further upstream [19].

Because of the differential effectiveness of downstream interventions, and an understanding of the social origins of health and health disparities, workshop attendees contended that there is a need for greater focus on upstream interventions in computing. However, it was noted that different "upstream" factors may be relevant to different marginalized groups. For example, stigmatization is a fundamental cause of health disparities among LGBT people [20, 21], whereas residential segregation is a fundamental cause of health disparities among African Americans [22, 23]. Accordingly, different intervention foci and strategies may be needed to influence social hierarchies depending on the group that is targeted in an intervention.

Workshop participants discussed the key capabilities of technology which can facilitate upstream interventions. Such technologies incorporate and allow for: (1) social coordination; (2) communication mediation; (3) optimizing resource distribution; (4) framing and supporting decisions; (5) educating; and (6) improving access to information. Based on this typology, attendees identified several possible types of interventions about which more research is needed; these are listed as open research questions below.

- How can technologies be used to reduce complexities for individual patients by redesigning work-flows for health maintenance and disease management?
- How can the (re-)design of technology-mediated interactions between dyads and small groups help to reduce stigmatization and discrimination?
- How can technologies enable and optimize resource pooling (e.g., transportation, food) in marginalized networks and communities?
- How can technologies be designed and implemented to facilitate collective resilience in marginalized communities?
- How can technologies catalyze self-organization around health in marginalized communities?
- What tools and data can help administrative and elected decision makers to positively influence health disparity-related policy decisions?
- How does technology contribute to unequal social determinants of health (SDOH) disparities and health disparities [24]? What approaches can mitigate these effects?
- To what extent can individual-level interventions have aggregated impacts on meso-level or macro-level factors?
- How can technologies be designed as multi-level interventions that simultaneously target individuals and meso- or macro-level factors?

## 2.2 Equity-Centered Intervention Uptake and Study Recruitment

**Researchers must recruit marginalized groups as research participants to advance sociotechical interventions for health disparities.** Workshop participants highlighted significant challenges with engaging some marginalized populations with technology — especially with respect to recruitment for studies and the uptake of technologies once they are made available to marginalized populations.

Workshop participants noted that many studies of health-focused socio-technical interventions are marked by selection bias due to recruitment challenges. In particular, it may be difficult to recruit racial/ethnic minorities — as shown in a Gibbons et al. systematic review [25] — as well as people of lower socioeconomic statuses (SES), and LGBT people. However, **it is difficult to fully assess representation in research on sociotechnical interventions due to widespread non-reporting of relevant demographic characteristics** (race, ethnicity, socio-economic status, language spoken, education level) in published computing research [9].

With regard to recruitment of research subjects, workshop participants identified the "standard" strategies that are commonly used to recruit study participants, such as: offering incentives; contacting prospects via phone, email, or mail; established participant pools; posting flyers; advertisements; crowdsourcing-based recruitment (e.g., via Amazon Mechanical Turk); and through healthcare organizations. **Participants noted, however that these approaches often fail to reach groups that experience health disparities, suggesting that current methods favor people who are demographically advantaged and already engaged in their health and health care.** Study eligibility requirements may also favor health-engaged people; indeed, some studies have required that patients be motivated to change their health behavior to be eligible for the research.

When discussing recruitment and uptake challenges, workshop participants repeatedly mentioned **the role of trust in study participation and technology use.** Indeed, a large body of research has shown that mistrust is partly due to previous mistreatment by medical researchers [26] and experiences of racism in health care [27]. Lower trust in technology may also play a role in the uptake of sociotechnical interventions, as trust is an antecedent to technology adoption and use [28-30].

Community-based participatory research (CBPR), an approach emerging from the public health field, has been successfully applied in many observational and interventional studies, including those using sociotechnical interventions [31]. CBPR methods can assist in the recruitment of marginalized groups through community organization partners. Furthermore, potentially with the help of CBPR, workshop participants recommended **framing the intervention around something that marginalized groups truly care about.** Workshop participants also advocated **recruitment strategies which avoid self-selection bias where possible, and which recognize and mitigate gatekeeper bias** (e.g., recruitment by a clinician) in participant selection. It is also important for researcher to understand and address reasons that potential participants may have for refusing study participation, such as transportation or child care.

Given the aforementioned challenges and progress, workshop participants saw a need for research specifically to evaluate methods of recruitment and promotion of technology uptake, and to develop and test novel methods. Workshop participants identified the following open questions in need of further research:

◗ How can theory inform efforts to enhance recruitment and uptake in marginalized groups?

◗ Why do marginalized populations enroll in studies at lower rates? How can we address these issues?

◗ Do "standard" recruitment strategies work for marginalized populations? For whom do they work? For what health issues/behaviors? When do they work?

◗ What new uptake/recruitment strategies are possible to reach marginalized populations and are they effective?

◗ How can we promote uptake of socio-technical interventions among disinterested populations?

◗ What are the best methods for building trust between interventionists/researchers and marginalized populations?

◗ What kind of technological training and support is needed for interventions with marginalized populations? How do they differ for different groups?

◗ Can technologies (e.g., location tracking) effectively assist in uptake/recruitment efforts in marginalized populations?

Participants felt that there were opportunities to learn from marketing researchers regarding strategies for selling products and community organizations that conduct outreach.

## 2.3 Equity-Centered Engagement/Adherence and Study Retention

Workshop attendees discussed **difficulties with both differential engagement/adherence with socio-technical interventions and study retention**, which was defined as a research subject continuing in the study until the last data collection point. Workshop participants also noted that, in field studies of sociotechnical interventions, it may be the case that **subjects engage with and adhere to an intervention while still dropping out of a study.**

Workshop attendees identified **challenges in defining active engagement** with sociotechnical interventions since the term "engagement" is not well defined. Some researchers defined engagement as usage, and others as more of a subjective experience. Subjective engagement has been linked to ongoing use of sociotechnical interventions in health [32]. Despite this conceptual distinction, **much research has focused on engagement operationalized as intervention usage levels.**

When operationalized as usage, a number of engagement-focused studies have shown that people with less formal education (an indicator of SES) use sociotechnical interventions less than those with more education, regardless of the intervention's level of structure [33-40]. Similarly, study retention, focused on completion of all points of data collection, is characterized by lower completion among those with less formal education. At the same time, **published papers typically do not report on dropout rates and the demographics of non-users, less-engaged users, and study dropouts.**

Workshop participants highlighted the importance of using technology design strategies for sociotechnical interventions that can assist in reaching those who most need them. **Participants' successful experiences supported a process involving needs assessment, participatory technology design, and community partnerships.** These partnerships were most successful when faithful to the principles of CBPR, including **equity** between academic and community partners. As Cortés found in urban communities [41], such models may also align with the expectations of marginalized groups for research involvement. Community involvement in developing strategies for promoting intervention engagement was also believed valuable.

When planning evaluation studies, participants advocated an experimental design that includes a specific protocol for engagement and retention. They also recommended **closely monitoring system engagement and intervening quickly to re-engage participants if necessary.** It was also thought helpful to predict when disengagement might happen, and proactively use strategies for re-engagement.

Given the aforementioned challenges, workshop participants saw a need for research specifically to develop and test existing and emerging methods of engagement and retention for marginalized groups. Workshop participants identified the following open questions in need of further research:

- What is the meaning of engagement with sociotechnical interventions from the point of view of different marginalized groups?

- What current and new engagement/retention strategies are possible to reach marginalized populations? When are they effective? How are they effective? What are their costs and benefits?

- How and when should we re-engage marginalized people who have ceased to use sociotechnical interventions or dropped out of evaluation studies?

- For sociotechnical interventions that engage groups or networks, how should we measure group dynamics to assess engagement?

- What ethical frameworks apply to the engagement/retention of marginalized participants? What limits should researchers observe in encouraging engagement/retention?

Workshop participants identified some needed resources, including **increased multidisciplinary research on problems of recruitment/uptake and retention/engagement.** Workshop participants also believed that **research funding opportunities should allow for more resources to be devoted to recruitment and retention** and that ethics boards should be educated to understand that methods of recruitment and retention require iteration and refinement over time. Finally, workshop attendees identified a need for a mechanism for sharing effective recruitment and retention strategies with different populations (e.g., advertising methods and keywords).

## 3. Sociotechnical Black Boxes

**When we develop interventions and applications that change people's behaviors or outcomes** — especially health related upstream or downstream interventions — **it is not always clear what part of the intervention, technology, design, community engagement, sociotechnical theory, or person's current mindset enabled that change.** Likewise, unless explicitly studied, it is unclear how long that change will be sustained. We use the term "black box" to highlight these ambiguities.

**Although improved health outcomes as a result of a sociotechnical intervention are exciting, they are not enough — we must also understand the mechanisms behind the change and potential "side effects"** so that we can reproduce the changes and continue improving on them. We outline the topics discussed by workshop participants and indicate the interplay between sociotechnical systems and sociobehavioral theory (if used) in Figure 2. As Figure 2 shows, the top three factors which we need to understand to "open" black boxes are (1) identifying how participants are involved in the intervention design; (2) understanding data quality from a hardware, software, and human perspectives; and (3) identifying the appropriate dosing of intervention use. Discussed less here, but as important, is integration with theory. These factors are important to understand independent of research aims — whether they are informative (where they describe what is happening) or actionable (that lead to design or adoption of new interventions).

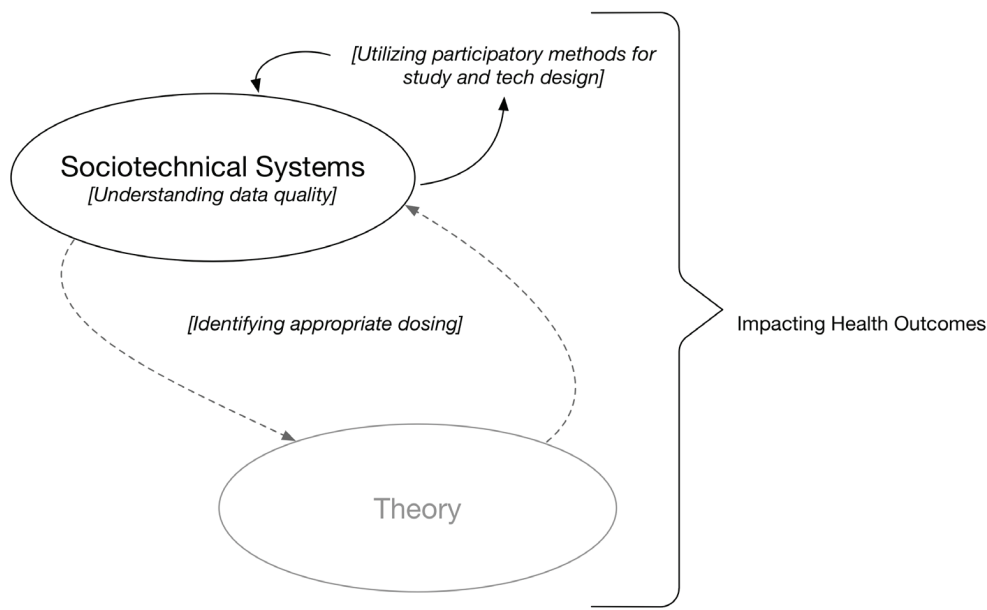

*Figure 2. Factors to consider when exploring sociotechnical black boxes*

## 3.1 Participatory Methods for Study and Technology Design

Workshop participants expanded on defining a science of engagement and retention (section 2.2) by discussing how to integrate marginalized people into the study and technology design processes. Researchers acknowledged that existing tools and methods used for research have implied criteria for success (e.g., significant differences; usability), **thus tools, methods, and criteria for success are conflated.** Researchers **mitigated these effects by actively collaborating with the community where the study was being completed** to investigate **real-world problems** which lead to broader, more contextualized measurements that could benefit the community, increase researcher's broader impacts, and strengthen their community collaborations. If **research communities do not support addressing validated real-world problems**, researchers run the risk of **solving the wrong (or artificially easier) problems** and not having a clear **sense of dynamic risks and unintended consequences.**

Researchers reported many ways to collaborate with communities — from university-led centers or initiatives in communities (e.g., a center for rural engagement,[1] community-centers in assisted housing neighborhoods[2]) to including community members as part of the research team [31] to remotely creating a community via social media [42]. Likewise, methods to involve community members varied — from including community members in research ideation to data collection, analysis, and dissemination. Researchers, who include community members in research, had to balance multiple tensions — especially when considering what is valued in scholarly research versus what is most beneficial to the community. A continued effort by **funders to encourage and support true broader impacts in research** will assist researchers in addressing this tension.

Researchers also discussed how to engage community members in the research process when some parts of research (e.g., research ethics training to deliver novel interventions or evaluate data; dissemination for publication) is time-consuming and an academically-oriented burden. In these cases, researchers emphasized the need to **ensure research benefits** not only the researcher, but **the community**, and to **adequately compensate community members at an equitable level to researchers,** since they are part of the research team.

Although there are many participatory methods — from CBPR to Action Research [43] to participatory design [44] —

---

[1] https://rural.indiana.edu/

[2] http://www.denverbridgeproject.org/

researchers noted challenges specifically for marginalized groups that the research community needs to explore further:

◗ How do we build capacity for communities to meaningfully participate? How do we mitigate power issues that may be perceived when working with people of varying backgrounds and experiences?

◗ How do we evaluate the effectiveness of participatory methods with marginalized groups?

◗ How do we provide a safe space for not knowing, learning, iterating, and reporting failure in participatory design when working with marginalized groups?

Researchers acknowledged the immense value of working with marginalized groups. Participants reported that formally trained researcher team members who were marginalized community members ("community research liaisons") assisted with recruitment, retention, and trust formation. So although participatory methods to involve community members are important, it is more so **important to ensure there are funded programs** to encourage, mentor, advocate for, and **train marginalized groups to become formally trained practitioners and researchers** in these research areas. As part of this, there is a need for expanded opportunities to train and support community research liaisons, including facilitating dialogue between liaisons. Creating a more inclusive research community will improve, strengthen, and push innovation to benefit everyone [45].

## 3.2 Understanding Data Quality in Existing Systems

In order to identify how sociotechnical black boxes work, researchers must consider the quality of the data generated in a socio-technical system. Workshop participants approached data quality from two viewpoints — methodology and provenance. Methodology refers to a study's design and how it can impact data quality generated from participants, instruments, systems, and study components. Provenance refers to the quality of data streams that people, technology, and inferences (e.g., machine learning) generate.

Computing researchers often investigate novel interactions, technologies, and infrastructures by conducting pilot studies [46] — which are not always recognized by health-oriented fields because of their small size, short duration, or lack of statistical power. In addition, although there are computing researchers investigating how to assist marginalized groups in improving their health, most studies are fairly short and difficult to compare [9]. Researchers, however, **stressed the importance of pilot studies and their important role in helping to ensure that starting conditions for interventions are correct.**

This view is not unique to computing; indeed, public health researchers have **advocated for treating pilot studies as an integral part of the scientific process** [47]. In addition, computing and behavioral medicine researchers have encouraged their communities to better report on data and contributions to identify causal effects of behavior change [48]. When conducting larger studies, particularly to assess health outcomes, a challenge is the difficulty researchers encounter when recruiting and retaining marginalized participants; thus smaller study samples may remain common. Two promising ways to strengthen conclusions with smaller samples and thereby **overcome the research-to-practice gap** include **adopting models that iteratively design sociotechnical systems** that are eventually sustainable without researchers [49] and **modularizing sociotechnical systems into the bare components** to identify their effectiveness — even on smaller sample sizes (e.g., agile science [50]).

During study analysis, researchers are strongly encouraged to **consider the provenance of the data streams** that people, technology, and inferences generate. There is also a need to improve our ability to account for the impact of complex social relationships on data collection and use in some groups (e.g., parent-child; patient-provider) and for the impact of user characteristics (e.g., age, health literacy) and environments (e.g., rural vs. urban areas) on data quality [51, 52].

When workshop participants discussed all of the ways in which researchers can collect data — from instruments to data streams — participants raised questions about **how much data to collect** in a given study. This is important in relation to both user burden and future-proofing the set of measures in the event of novel research questions which may emerge over time. In general, lower user burden is associated with more successful data collection — a phenomenon which may be amplified with marginalized groups, such as people with low SES [51]. This concern would tend to favor collection of less data. Indeed, for a given goal or set of questions, a small set of measurements, taken infrequently and with a focus on trends or absolute accuracy, may be all that is required. However, future-proofing may favor collection of more data: as research questions or patient goals change, researchers or individuals may wish they had collected more data to facilitate re-use for these new questions or goals. At present, researchers may lean towards collecting more data to future-proof, but this may increase user burden. Furthermore, these data may not all ultimately be used and there may be some privacy and security risks to having these data in a researchers' possession. Participants thus felt that there was a need for **more researcher attention to how much data to collect, for what purposes, and for whose benefit.** One suggestion that emerged from the workshop was to conduct more studies in which measurements themselves are studied by varying what we measure, how we measure it, and on whom the measurement is performed.

More specifically, researchers are encouraged to investigate:

◗ How do we identify and report on data provenance — especially in marginalized groups?

◗ How can we better communicate algorithmic abstractions so that all stakeholders understand the limitations of the data provided?

◗ How do we understand the implications of the technical measures and data streams that we collect? What are the impacts of the characteristics of marginalized users and their living or working environments on data quality from sensors, location tracking, connected medical devices and self-reports?

◗ How can we better understand what we measure — especially unobtrusive measures gathered from sociotechnical systems that may vary by context?

One possible unintended consequence of designing systems that improve the health of people with health disparities is that it could also benefit those with health-related advantages; thus we may continually raise the baseline of health and sustain health disparities or make the gap larger. Consequently, we need to **continually monitor data to assess community health baselines and health disparity gaps.**

### 3.3 Designing Dosing Schemes

We must consider the ideal or actual "dosage" of sociotechnical systems for their users. **Dosage refers to the frequency and intensity of user experience with sociotechnical systems —** including use of technological features, interactions between system users, and exposure to theory-informed behavior change techniques. As part of this, the context in which a technology is employed is relevant to dosing because exposure to elements of the social and community context may create enabling or constraining conditions for intervention effectiveness; in this sense, a certain "dose" of community walkability and safety from crime may make it more possible to benefit from physical activity more often. Or, a certain amount of social support — whether included in the design of a technology or not — may be needed to benefit from a mental health intervention. Within this framework, researchers encouraged more research regarding:

◗ Understanding how often a dose of a sociotechnical system should be given (e.g., daily, as needed, in a structured program of a pre-specified length) and what mechanism to use for to administer a "dose" — which will change depending on one's context and experiences.

◗ Evaluating and reporting on the burden-engagement trade-off of different dosage schemes and of different parts of sociotechnical systems (e.g., participant burden using the system remotely or in-person; research burden managing the data streams).

◗ Investigating missed doses because non-use or inactivity within each dose does not necessarily imply that the intervention has failed or that change is complete.

◗ Understanding what mix of dosing of technological and non-technological elements are needed to achieve a given outcome; for example, ways in which community or social network characteristics may moderate intervention effectiveness.

◗ What role can novel dosing schemes such as pulsed, decreasing or event-based schemes play in sociotechnical interventions for marginalized groups?

Once dosing schemes are established, **more research must be done to investigate their reproducibility** in new contexts. Reproducibility is especially difficult in sociotechnical intervention research because norms and technology move quickly while infrastructure availability and technology adoption among marginalized groups tends to lag behind those with greater resources. **We were encouraged by federal initiatives,** such as the FCC-NCI Broadband Cancer Collaboration,[3] **to enhance infrastructure for groups that experience health disparities.**

**We must be able to model, interpret, and communicate about dosing schemes to better understand how they should be designed, and when different options should be used, for which group, and in what context.** For example, given their implications for engagement, when should researchers designing self-monitoring interventions utilize sensing versus self-report? How much in-person, virtual, or asynchronous communication with others (e.g., healthcare providers, patient peers) is needed to have an intervention effect? In addition, **we must be able to effectively communicate dosing information to researchers in other disciplines and lay community members.** Dissemination must be more than writing up results and throwing them over the wall — **we must actively pursue new mediums to communicate findings** to people with various backgrounds. Researchers could envision tools to assist researchers and practitioners in deciding on the optimal dose of intervention components for specific health outcomes in specific populations. In addition, **researchers should embrace succinct visual and multimedia communication techniques to justify dosing to non-experts and lay populations** all the while emphasizing how the study design meets their community-based research goals.

## 4. Sociotechnical Systems to Inform Theory

One of the biggest challenges of developing sociotechnical health interventions is that although there are many sociobehavioral theories available [53], they are often dated because they do not account for new types of sociotechnical systems [10, 50] and are not necessarily representative of marginalized populations. Indeed, workshop participants could only identify two health behavior theories or models that were developed specifically with marginalized populations — the Theory of Positive Deviance [54, 55] and the Reserve Capacity Model [54]. Therefore, often used, but dated sociobehavioral theories may have limited predictive power for marginalized groups in a digital age. Additionally, new types of data (e.g., continuous sensor monitoring, social media streams) cannot be automatically used to validate theories that were created before this type of continuous data was available.

Despite the aforementioned shortcomings of existing sociobehavioral theories, **researchers are encouraged to select theories and define theoretical constructs** for use in study design; designing interactions within the sociotechnical intervention itself; or analysis of data [10]. By identifying an explicit theory and related constructs, researchers articulate their assumptions and provide a record to explain how an intervention worked. Although choosing an explicit theory forces the research team to make assumptions, not doing so may mean relying on implicit theory to which all research team members may not have agreed. **When defining a theory, researchers should also cite the origin of the theory and how, if at all, the target populations of the sociotechnical intervention differ from the population in which the theory was developed.**

### 4.1 Building Better Theories: New Opportunities

Traditionally, use of theory has been one-way — an individual or community's behavior informs an abstracted understanding of what is happening to create a theory (A→B). Theories can be extended or new theories developed as prior ones no longer express what is observed (black arrows in Figure 3). However, this approach is insufficient when technology is taken into account because behavior is dynamic and changes depending on context and time; moreover, technologies can both create and capture variance. In addition, researchers attempt to map abstracted sociobehavioral theory onto concrete technology interactions (blue arrow in Figure 3). A common example is displaying one's history of past actions (e.g., a food or activity log) as evidence of past performance as part of feedback provision in an intervention, then logging how many times a user accesses the history screen as evidence on a participant's reflection on past history. However, it is often unclear how well such measures truly map onto theoretical constructs. Accordingly, researchers must identify ways to map sociobehavioral theory appropriately onto sociotechnical systems and evaluate the scope of technology mapping in theoretical constructs.

---
[3] https://www.fcc.gov/health/cancer

To facilitate theory development, participants identified a need for new methodologies that can learn both new "hypotheses" and construct new theories or extend current theories using data, and adapt as more data and data types emerge. Moreover, these theories need to be specifically developed to reflect the experiences of marginalized users and the under-resourced contexts in which many are more likely to reside. These theories also need to explicitly address the meso- and macro-level factors from which disparities emerge (see section 2.1).

A promising area of interdisciplinary research for theory development is just-in-time adaptive systems [56, 57]. In this approach, depending on one's dynamic behavior and context, the sociobehavioral model is updated along with the sociotechnical systems' interactions with the world, thus creating a dynamic system (green, dashed arrows in Figure 3) that can adapt and provide relevant information to the user and research teams. In adaptive systems researchers must address many challenges:

- What data streams can best inform dynamic theories that address the social origins of health and health disparities?

- How can researchers utilize triangulation to identify better data streams and come closer to a ground truth understanding for marginalized groups that are not well represented in current literature?

- How do we build on sociobehavioral theories that capitalize on the dense contextual and behavioral data that sociotechnical systems can collect? How can such theories then improve behavioral prediction and dissemination to policy makers to improve social determinants of health?

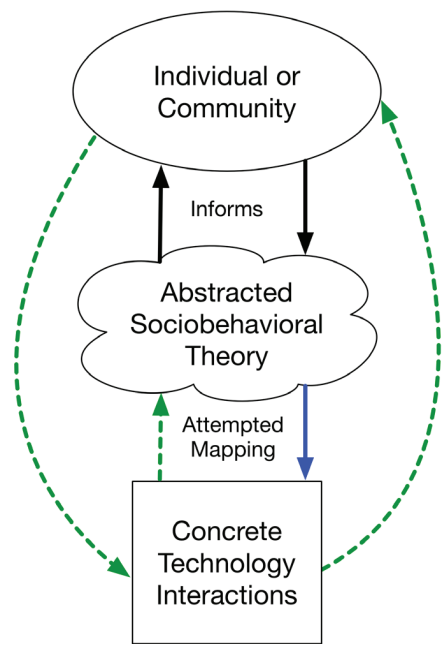

*Figure 3. Example of relationships between individuals/communities, sociobehavioral theory, and technology interactions*

- How do we iteratively develop and validate dynamic and personalized theories in marginalized populations?

- How do we decide if a theoretical construct is not useful for a specific population, setting, or technology?

- How do we identify blind spots that theory can introduce? In the face of potential blind spots, how do we know what we are not measuring, and for whom — especially in marginalized groups?

## 4.2 Tailoring and Optimization of Sociotechnical Systems

Health communication researchers differentiate targeting an intervention for a specific population from tailoring it based on the needs of an individual. Targeting means that an intervention's design would be created specifically for a group's needs — for example, a hypertension management application for African American women. Tailoring means that the needs of one person are assessed and the system's output personalized based on those needs [58] — for example, a hypertension management application for a specific African American woman.

For marginalized groups in particular, targeting and/or tailoring are recommended since existing sociotechnical systems may, as mentioned earlier, reflect the biases, worldviews, experiences, and assumptions of their (typically) more advantaged designers. To facilitate targeting or tailoring, **researchers are encouraged to cite the instruments and data streams they used to understand the rich context and lives of the populations with which they work. They should discuss how these data were used to target or tailor the intervention.** We also strongly encourage researchers to reflect on their interventions and share what data they wish they had and did not collect; this may help the research community to identify a core set of data points based on populations and context.

Tailoring applications in particular populations creates a wealth of challenges for interdisciplinary researchers, such as:

- How do we collect all of the data needed for tailoring without unduly burdening participants or violating their privacy — especially in marginalized groups that have previously been exploited for research gain?

- How do we make sense of all of the data streams to ensure they are providing an accurate picture of users' contexts, lives, and cultures? For instance, one may portray themselves differently in everyday life versus what they share via social media. How do we prioritize data streams in cases in which a marginalized person (e.g., a transgender or gender non-binary person) may need to protect their identities for safety?

- How do we develop adaptive interventions that can dynamically adapt to life changes while creating a consistent interaction with the system for participants' comfort?

- How do we design adaptive visualizations that people can understand — especially with varying literacy, numeracy, and language proficiencies — and act on them?

- How do we evaluate socio-technical systems when they are so tailored that each participant has an individualized experience? How do we replicate studies when experiences are individualized?

A further caveat for intervention targeting and tailoring is that the research community must have **checks and balances in place to ensure the adaptive algorithms are providing meaningful information without unintended consequences** such as discriminatory practices [7, 59].

## 5. Multidimensional Evaluation to Reduce Health Disparities at the Population Level

The multidisciplinary workshop provided participants with ample time to share experiences designing, implementing, and evaluating studies at various levels of granularity — from individuals to families to communities and, finally, to the population level. Workshop participants discussed the need to measure multiple dimensions — contextual, structural, and social determinants of health (e.g., Figure 1, page 5) - to better evaluate changes with respect to health disparities.

## 5.1 Improving Measurement and Methods for Multidimensional Evaluation

For each project, **researchers and intervention designers must decide upon the "right" set of factors to measure — balancing trade-offs in practicality, comprehensiveness, strategic value, and risk.** Researchers may select measures to produce novel findings that are informative (e.g., describe what is happening or why) and/or actionable (e.g., lead to design or adoption of new interventions). Researchers may also adapt measures from a well-understood intervention to ensure that it works as they scale it up or roll it out to practice. A **caveat is that researchers should not limit themselves to what can easily be measured.** Indeed, "real world" success may be especially difficult to assess.

### *5.1.1. Measurement*

With regard to **what to measure, it is important to measure health equity-relevant outcomes,** which may assess different types of equity (e.g., healthcare access, financial access, health behaviors, health literacy, healthcare quality, health-related outcomes). It is also important to know how well a specific intervention actually implemented a theory, and to know what parts of an intervention generate treatment effects. To this end, **researchers must evaluate the quality of prototype systems, intervention usage, and theoretical mechanisms of action (mediators) at different stages of the posited causal pathways**, as well as the outcomes that the intervention is intended to influence. **To show a change, these measures must be assessed longitudinally**. Furthermore, researchers must understand and document for whom the intervention has an effect, and in what circumstances (moderators).

Measure selection is complicated in health disparity contexts because health disparities emerge from social conditions at multiple levels (macro, meso, micro). It may be necessary to develop upstream interventions or synchronous multilevel interventions (see Section 2 and Figure 1 on page 5). Furthermore, an intervention operating at any level may have effects at another level (e.g., policies to provide women with access to education may increase their power in intimate relationships, thus empowering them to insist upon condom use and reduce their HIV risk).

An open area of research is to identify what level of analysis should be used in measurement.

**Theory can guide selection of these measurements in evaluation of interventions, however, it is difficult to ground evaluations of upstream and/or multilevel interventions** in theory due to the lack of maturity in available frameworks. For example, similar to sociobehavioral theories, upstream and multi-level theories lack dynamism and do not necessarily account for bi-directional relationships between different levels of social conditions or between social conditions and individual characteristics and behavior. Mechanisms and drivers of change may also be unclear. The lack of theoretical guidance makes evaluation challenging in this context; thus there is a **need to improve upstream and multi-level theories to improve measurement, and vice versa.**

Participants also noted that **measures used in disparity research tend to focus on deficits and barriers, rather than resilience and facilitators.** Attendees advocated measurement of a broader range of phenomena in our studies, with an emphasis on developing a fuller picture of marginalized groups and individuals. Workshop participants **encouraged researchers to share what they measure — including challenges and strengths**, while also working towards **a widely accepted set of metrics to assess intervention impact at different levels.**

Despite the plethora of sociotechnical measurement options, there is a **need for expanded technical capabilities to measure a wider range of equity-relevant characteristics**, such as culture and patient goals. Furthermore, there is an opportunity for developing better measurement methods that leverage new data sources, such as characterizing digital phenotypes based on patterns of user interactions with interventions, and trajectories of usage over time. **A recurrent challenge, however, is that patient-centered measures are often individualized, not standardized, which makes reproducibility difficult.** This is especially true in marginalized groups that are less studied; thus, researchers have less baseline and longitudinal data.

The workshop surfaced the following open questions in need of further research in planning multidimensional evaluation measurement:

◗ What is the relative value of different measures for improving predictions? What are the relative costs and benefits of using different measures?

◗ How can health-equity-relevant phenomena such as resilience, culture, context, and patient-centered goals be more effectively measured? How can we balance individualization and standardization in creating these measures?

◗ What patterns of user interaction with interventions exist? How do they vary for different marginalized groups? How do they change over time? How do these patterns influence intervention effectiveness, if at all? What can they tell us about when interventions should end?

◗ How can we assess the quality of a theory's operationalization within a socio-technical intervention?

◗ To improve extant theories and models, how can we better measure mechanisms of action (mediators) and identify groups/settings in which interventions are effective (moderators)? How can we measure multi-level outcomes within individual studies?

### 5.1.2 Methods

Like technological interventions, upstream and multi-level interventions may also fail to fit well into the existing Randomized Controlled Trial (RCT) paradigm. Implementation may be "nonlinear, iterative, and adaptive" [60], leading to differences in interventions over time or across sites. Evaluations may be complicated because of a need to collect data about effects at different levels (e.g., both community norm change and individual behavior). Furthermore, it is not always clear what the "active ingredients" of interventions are, and how those ingredients interact — key requirements for understanding generalizability and translation. Workshop participants **identified a need for using more varied existing study designs to find the intervention components that do or don't work.** Relatedly, there is a need to further apply and develop new adaptive trial methods such as multiphase optimization strategy (MOST) and sequential multiple assignment randomized trial (SMART) [61]. **Optimization and tailoring criteria specifically related to marginalized groups and the social factors that drive disparities would increase the applicability of these evolving methods to disparity research.**

Participants also noted that **follow-up, including long after an intervention, is a critical missing piece in prior research.** Such longitudinal follow-up may prove more difficult to conduct in low-SES groups due to less home ownership and, potentially, more contingent and precarious employment. Accordingly, there is a need for further study of **methods for retaining these groups in research** (see Section 2.2).

Participants also highlighted the **need for greater understanding regarding implementation of successful interventions in new settings, and with new marginalized groups.** A related issue of concern is the potential for effect modification based on contextual factors for research with marginalized groups. Workshop participants thus voiced a need for **cross-cutting studies with multiple comparisons across different marginalized groups.** Translation also involves potential re-use of measures, data, and technical frameworks. Therefore, participants advocated **additional research on how to best facilitate re-use** [49, 50]. In a health disparity context, it may be necessary to facilitate re-use for older technologies that are more widely used in low-SES groups, such as interactive voice response, SMS, and 2-G telephones [51].

In addition to collecting data for analysis, data can be used to design predictive models that provide opportunities to intervene to amplify the treatment effect (outcome), or mitigate intervention risks. However, these models are only useful if we can characterize highly complex behavior systems through interactions between sub-systems and dynamics over time. Workshop participants challenged the scientific community to consider:

◗ How can we better measure intervention effects across micro, meso, and macro levels? How can we effectively account for interactions between different levels of outcomes?

◗ How can criteria specifically related to marginalized groups and the social factors that drive disparities be incorporated into sociotechnical interventions as optimization and/or tailoring criteria?

◗ How can we best facilitate re-use of measures, data and technical frameworks specifically for marginalized populations?

◗ How can we characterize complex systems in predictive models regarding disparities?

## 5.2 Assessing Equity Impacts and Unintended Consequences

Since marginalized groups are not often included in research (see Section 2.1), technology models may make assumptions that are not valid for them. Consequently, marginalized groups may miss potential benefits of an intervention — potentially worsening disparities. Workshop attendees argued for a **research process which continuously questions who may be left out in the design, implementation, and evaluation of socio-technical interventions to monitor the potential unintended consequences.**

To begin, there is a need for any intervention to specifically measure intervention outcomes. Furthermore, there is a need for **expanded effort to assess the equity impacts of both existing and emerging "universal" informatics interventions that are intended for all, rather than just a disparity population.** For prospective studies of new interventions, it is important for outcome evaluations to include planned heterogeneity of treatment effect (HTE) analyses. This requires recruitment of diverse samples, possibly oversampling from marginalized groups to ensure statistical power for such analyses. Such analyses may involve planned moderation, stratified, or subgroup analyses. Workshop participants noted that an area of ambiguity in such analyses concerned how to evaluate intersectionality and overlapping disparities in intervention evaluation participants. It is also important for intervention studies to include a qualitative component to assess possible unintended consequences in relation to health equity.

Additionally, for interventions that already exist (e.g., self-tracking of eating, exercise, or symptoms) there is a **need for equity-focused, interdisciplinary systematic reviews that aggregate individual patient data across studies.** One benefit of this approach would be the ability to include marginalized participants from studies which originally lacked statistical power for HTE analyses.

Open questions include:

◗ What, if any, "universal" intervention types and designs perform better in marginalized groups than the general population?

◗ How can we effectively evaluate the effects of interventions in situations of overlapping disparities?

◗ What are the equity-related consequences of previous socio-technical interventions in health?

◗ What are the health equity impacts of consumer health technologies that are now in wide usage (e.g., self-tracking, patient portals)? What are the population-level impacts of any identified inequities?

For progress on unintended consequences, participants identified a **need to create a research culture in which learning — including failures — are embraced.** As part of this, workshop attendees wished to see the development of venues in which unintended consequences can be openly discussed, including panels and workshops to discuss equity-related lessons learned in projects. Furthermore, participants wanted to **encourage the development of scholarship on unintended consequences**, including **explicit sections on unintended consequences in publications,** complete papers on unintended consequences, and systematic reviews of equity-related unintended consequences of socio-technical interventions in health.

Funders also have a role to play. Funders should support **mechanisms to adjust interventions as they are implemented, allowing for a more iterative approach to research.** In addition, there is a need for funding research on, and reporting about, **equity-related unintended consequences** as part of grant progress reports.

## 5.3 Ethics of Conducting Sociotechnical Research with Marginalized Groups

Workshop participants stressed that **ethics for sociotechnical interventions includes the traditional medical ethical principles of beneficence, autonomy, justice, and non-maleficence.** The following ethical principles also govern our activities as sociotechnical researchers in health: competence and scientific validity; responsibility (to others and society); respect for potential and enrolled participants in a study; respect for privacy; honoring of confidentiality; equity (e.g., in research subject selection); honesty and trustworthiness; favorable risk-benefit ratio; informed consent; due process; transparency in data collection and usage; and atonement. Relevant principles acknowledge that harm may include physical, financial, social, and psychological/emotional injuries. While these principles are widely supported, **their application in a digital world, presents many uncertainties in need of further investigation.**

### *5.3.1 Responsibility*

Workshop participants emphasized the view that **evaluation of sociotechnical interventions must have a clear vision of success** — and one that is not fully bound by the methods and tools that we use in our research. Unfortunately, research and academic priorities are frequently misaligned with marginalized groups' priorities. This creates a challenge in establishing the direction of research, including whether research questions are generated by the needs and concerns of the community or whether researchers are trying to solve "problems" that only exist in theory, but are not critical to addressing the key causes of disparities. Consequently, some participants advocated an **emphasis on "real world" success**, or impacts on people's lives outside of the academy. To define real world success, participants advocated dialogue from diverse stakeholders. As a starting place, participants noted that success would likely involve: (1) people in the greatest need receiving the support they require; and (2) creating circumstances that enable more people and groups to help themselves, and then share their learning and insights with others.

Participants noted the challenge that, as sociotechnical interventionists, we may be **overly-focused upon technical capabilities and solutions, making it difficult to fully understand resources that marginalized groups currently have, what they genuinely want and need.** For example, **technology may not be the right solution for a given problem, and there are circumstances in which technology may make things worse for marginalized groups** (e.g., self-tracking of eating linked to worsening of an eating disorder [62]). Accordingly, researchers should **avoid being overly prescriptive and be open to negotiation as to possible activities and directions.**

### *5.3.2 Return of Results*

A significant ethical issue discussed by participants concerned the **return of research results to participants, which can viewed as a way to enact respect for participants** [63]. This issue covers both the return of individual study results [63] and aggregate study results, as in research conducted within a low-income neighborhood. Workshop participants contended that, in the face of challenges common in marginalized groups, such as health literacy, there is a critical need to **investigate novel methods for disseminating research findings.** Promising techniques may include storytelling and accessible visualizations. At the same time, there is a need for plain language and clear communication in the area of informed consent.

### *5.3.3 Risk-Benefit Ratio*

Sociotechnical interventions typically involve the collection of a great deal of data about their users/study participants. Studies may involve the deployment of third-party platforms that are not owned and controlled by researchers (e.g., Fitbit, Apple ResearchKit). In such cases, it may not be entirely clear whether patients can opt out of data collection, whether stored data are de-identified, where data are held, and whether researchers or participants can delete said data. Furthermore, research participants' privacy and security will be subject to the platform's practices, and data may be vulnerable to sharing with unknown additional parties. Such difficulties may be particularly risky for study participants with stigmatized identities or conditions, such as transgender people, people with HIV/AIDS or mental health conditions

> **For a given intervention researchers should consider asking themselves:**
> ◗ Who can easily access the intervention? What barriers might marginalized groups face in trying to access it?
> ◗ What are the potential unintended consequences of my research, particularly for marginalized populations?
> ◗ What is the impact of the intervention on health equity?
> ◗ Was there heterogeneity of treatment effect? If so, for whom, and how much was the effect size difference?
> ◗ Who engaged with the intervention? What was the impact of any differential engagement levels on intervention effects?
> ◗ Is there differential dropout or abandonment of the intervention in marginalized groups?
> ◗ What would the outcomes be in marginalized groups who were not reached?

Box 2: What should intervention researchers consider when designing interventions?

[64]. More research must be done to create models that researchers in any discipline can use to assess risks of study participation for marginalized groups.

Furthermore, sociotechnical interventions are increasingly gathering and using contextual data, which includes participant demographics, place of residence, geolocation, call and text logs, social networks, patterns of technology usage, keystrokes, and biometrics. Some just-in-time adaptive interventions focus on delivering support and information when needed, based on data gathered from geolocation and sensing technologies [64-67]. For instance, contextual data such as call and text logs and geolocation have been used to sense a person's mental well-being [68, 69]. Furthermore, healthcare organizations are increasingly collecting social determinants of health data and incorporating these data into electronic health records [70, 71]. More and more clinical prediction algorithms are incorporating contextual variables, such as whether a patient lives in a high-poverty neighborhood, into their models [72, 73]. The ethical implications of gathering this growing amount of data are unclear, and they may represent greater risks for marginalized individuals.

### 5.3.4 Informed Consent

With an increasing amount of data collected about people in their everyday lives and as part of studies, questions regarding informed consent become increasingly complicated. Secondary use of data may involve data, such as digital traces from such a large group of people, that make traditional methods of obtaining informed consent become infeasible. Furthermore, if study participants share accounts and technology, then the definition of a "participant" may have to be further specified.

People may also find themselves participating in research, especially social media platforms [74], without intending to do so. To illustrate, patients may unwittingly find their contributions in online patient communities used for research, and details on secondary usage are often buried in terms of use and privacy policies that users infrequently review [75]. Contextual data may involve others who have not consented to be part of a study, such as Facebook friends or conversation partners on a smartphone. In each of these cases, there is a potential impact on the autonomy of all study participants, but the impact may be greater in individuals with low health literacy or people who experience stigma and discrimination in their daily lives, such as those with substance use disorders.

With the aforementioned ethical issues in mind, participants identified the following open questions:

◗ How can we include more diverse perspectives in discussions of research ethics for sociotechnical health interventions?

◗ How do we ensure the data we give back to study participants and groups is actionable, meaningful, and understandable? What is the impact of returning data to participants upon their willingness to engage in research in the future?

◗ What are the risks of capturing contextual data for marginalized individuals and groups? How can these risks be mitigated or managed?

◗ What are the ethical implications of use of contextual data to tailor interventions, diagnose health conditions, and identify health risks?

◗ What models of informed consent can be used for large-scale secondary use? How can informed consent models take unwitting research participants into account, if they constitute a participants' context?

Workshop attendees also identified a need for safe and encouraging spaces for discussing ethical dilemmas, sharing experiences and exemplars, and garnering resources. Workshop participants saw value in **developing a system to assist researchers, participants and companies in assessing ethical issues, including tools for proactive risk assessment and for balancing the inherent trade-offs in choices.** Another type of tool would assist study participants, researchers, and companies in developing privacy literacy. Mechanisms for reconciliation and remediation in the event of ethical breaches or harms were also desired.

## 6. Interdisciplinary Bridges

Many diverse populations are affected by health disparities; thus different, adaptable sociotechnical intervention approaches are necessary to help address the needs of individuals, communities, and populations. Currently, **researchers are largely developing separate approaches from scratch and in relative isolation or small interdisciplinary teams,** which makes it difficult to create scalable progress and larger real-world impacts. Without interventions that can scale up, our solutions are of reduced effectiveness, limited only to those who can afford them or happen to be in the right geographic area.

We recommend the development of a consortium or national centers to address health disparities with sociotechnical systems that creates a collaborative network of researchers, industry, providers, payers, and communities to aid in scaling sociotechnical interventions — similar to the NIH funded Center of Excellence for Mobile Sensor Data-to-Knowledge (MD2K). [4] **This consortium would act collectively to "raise all boats" by creating reusable components, sharing algorithms and data, developing approaches for transferability and robust partnerships, and developing the science of recruitment and retention of underserved populations in pilot and longitudinal studies.**

The consortium or coordinated national centers could have annual "themes" to drive collective action (e.g. "measuring stress") and teams could contribute measurement tools and data with respect to study design, recruitment, retention, sociobehavioral models, and dosing related to specific populations. **The consortium/national centers could put researchers into cohorts who are working with similar populations, dosing, or theoretical constructs** to build on each others' successes and failures and **improve translation of the research from pilot to community impact.** At the end of each year, the consortium/national centers would converge on a standard metric or tool that could then be broadly adopted. The consortium would also emphasize team science and **promote the next generation of interdisciplinary researchers in these areas by building a pipeline of underrepresented scholars and highly represented allies from undergraduate to early career researchers.** The consortium would need resources devoted to both research and sustaining community engagement by involving stakeholders throughout research. They would also incentivize data sharing and community engagement.

## 7. The Future of Sociotechnical Systems to Address Health Disparities

Studies funded by the National Institutes of Health have been formally investigating how to address health disparities for almost three decades [76]. However, health disparities persist — suggesting the need for fresh approaches. To that end, in this workshop report, we highlight computing, health informatics, behavioral medicine, and health disparities research challenges that cut across disciplines and federal funding agencies. We also stress the many opportunities that emerge from these challenges. They are summarized in the table below.

---

[4] https://md2k.org/

| Challenge | Opportunity |
| --- | --- |
| Marginalized groups are understudied because of difficulty with recruitment, retention, or trust issues. | Ensure researchers have resources to build and maintain community-based research collaborations.<br><br>Develop and evaluate methods of recruitment, technology uptake, and study retention for studies that work with marginalized communities. |
| Current sociotechnical interventions focus on "downstream" interventions where a participant manages a set of issues specific to themselves. Downstream interventions do not address the social origins of health disparities. | Support is needed to develop upstream and multi-level interventions to reduce health disparities by impacting community, social, economic, and political factors. |
| When we create sociotechnical interventions that have an impact on outcomes, it is not clear what part of the sociotechnical intervention initiated and maintained that change. | Encourage funding agencies to continue supporting broader impacts in research to ensure researcher are addressing issues that are important to communities.<br><br>Emphasize the need for pilot studies and iterative design to ensure initial conditions are correct.<br><br>Evaluate the "dose" of sociotechnical systems to better understand the frequency of use, as well as the dosing contexts and infrastructure support available. |
| Current behavioral theories and models often do not account for sociotechnical systems and are not representative of marginalized populations. | Document instruments, data streams, and mappings between sociotechnical systems and theories used.<br><br>Develop dynamic new theories that can account for future sociotechnical systems and capture the social contexts of marginalized populations. |
| Researchers must measure multiple dimensions of social determinants of health to evaluate impact at the population level, but there is a lack of dynamic theories, study designs, or metrics to capture the changing technological and contextual landscape of marginalized populations. | Create and document equity-relevant metrics that can capture appropriate levels of detail to contextualize user groups and interventions.<br><br>Develop, evaluate, share, and validate study designs and theories for interventions. |
| By designing to improve health disparities, researchers may introduce unintended consequences (e.g., everyone benefits and thus the disparities stay the same or worsen). | Establish research processes that check on what groups, data, or resources are unaccounted for and monitor unintended consequences.<br><br>Ensure data collection about unintended consequences.<br><br>Engender a research culture in which learning, sharing, and disclosing failures are encouraged. |
| Based on past treatment in research, some marginalized groups may have less trust in research. These trust issues are exacerbated when it is unclear how study participation or data access – especially in commodity products – is scoped. | Produce systems that assist researchers in identifying ethical issues and proactively assess risks with benefits. |
| Researchers in multiple disciplines are encountering similar issues in their research endeavors to address health disparities, but continue working in their disciplinary silos – sometimes reinventing each others' approaches or solving the same problems. | Develop a consortium or national centers to address health disparities that bring researchers from multiple disciplines together with partners to address the research to practice gap. |

We also encourage the **scientific enterprise to better align incentives** (e.g., funding, resources, tenure, publication) with **helping people — especially those who are marginalized.** Although there are alternative funding models that could be promising to encourage people to address health disparities (e.g., funding people and not projects [77]), we also acknowledge that with the dearth of underrepresented groups in research — especially computing — these models may not adequately support innovation in sociotechnical interventions for health disparity reduction. Workshop participants recognized a broader need to align the scientific enterprise with helping people. Specific to academic research, an easier mechanism for aligning incentives is to **add a fourth "impact" pillar for hiring, promotion, tenure, and merit reviews that goes beyond the traditional pillars of research, teaching, and service [78].**

## 8. References

[1] Kirchner L. When Discrimination Is Baked Into Algorithms. In: The Atlantic [Internet]. 6 Sep 2015 [cited 28 Jul 2018]. Available: https://www.theatlantic.com/business/archive/2015/09/discrimination-algorithms-disparate-impact/403969/

[2] Pearson J. How Big Data Could Discriminate. In: Motherboard [Internet]. 16 Sep 2014 [cited 28 Jul 2018]. Available: https://motherboard.vice.com/en_us/article/d73x7v/why-the-federal-trade-commission-thinks-big-data-could-be-discriminatory

[3] Lohr S. Facial Recognition Is Accurate, if You're a White Guy. The New York Times. 9 Feb 2018: B1.

[4] Cook KE. Marginalized populations. In: Given LM, editor. The SAGE Encyclopedia of Qualitative Research Methods. SAGE Publications; 2008. p. 296.

[5] Hankerson D, Marshall AR, Booker J, El Mimouni H, Walker I, Rode JA. Does Technology Have Race? Proceedings of the 2016 CHI Conference Extended Abstracts on Human Factors in Computing Systems - CHI EA '16. New York, New York, USA: ACM Press; 2016. pp. 473–486.

[6] Kirkpatrick K. Battling algorithmic bias: how do we ensure algorithms treat us fairly? Commun ACM. 2016;59: 16–17.

[7] Sweeney L. Discrimination in Online Ad Delivery. Queueing Syst. 2013;11: 10.

[8] Abbott J, MacLeod H, Nurain N, Essombe Ekobe G, Patil S. Local Standards for Anonymization Practices in Health, Wellness, Accessibility, and Aging Research at CHI. Proceedings of the 2019 CHI Conference on Human Factors in Computing Systems (CHI '19). New York: ACM; 2019. doi:10.1145/3290605.3300692

[9] Stowell E, Lyson MC, Saksono H, Wurth RC, Jimison H, Pavel M, et al. Designing and Evaluating mHealth Interventions for Vulnerable Populations: A Systematic Review. Proceedings of the 2018 CHI Conference on Human Factors in Computing Systems. ACM; 2018. p. 15.

[10] Hekler EB, Klasnja P, Froehlich JE, Buman MP. Mind the theoretical gap: interpreting, using, and developing behavioral theory in HCI research. Proceedings of the SIGCHI Conference on Human Factors in Computing Systems. ACM; 2013. pp. 3307–3316.

[11] Lai AM, Hsueh P-YS, Choi YK, Austin RR. Present and Future Trends in Consumer Health Informatics and Patient-Generated Health Data. Yearb Med Inform. 2017;26: 152–159.

[12] Huh J, Koola J, Contreras A, Castillo AK, Ruiz M, Tedone KG, et al. Consumer Health Informatics Adoption among Underserved Populations: Thinking beyond the Digital Divide. Yearb Med Inform. 2018;27: 146–155.

[13] Morton RL, Schlackow I, Mihaylova B, Staplin ND, Gray A, Cass A. The impact of social disadvantage in moderate-to-severe chronic kidney disease: an equity-focused systematic review. Nephrol Dial Transplant. 2016;31: 46–56.

[14] Lehne G, Bolte G. Impact of universal interventions on social inequalities in physical activity among older adults: an equity-focused systematic review. Int J Behav Nutr Phys Act. 2017;14: 20.

[15] Welch V, the PRISMA-Equity Bellagio group, Petticrew M, Petkovic J, Moher D, Waters E, et al. Extending the PRISMA statement to equity-focused systematic reviews (PRISMA-E 2012): explanation and elaboration. Int J Equity Health. 2015;14. doi:10.1186/s12939-015-0219-2

[16] Mynatt E, Hager GD, Kumar S, Lin M, Patel S, Stankovic J, et al. Research Opportunities and Visions for Smart

and Pervasive Health [Internet]. Computing Community Consortium; 2017 Jun. Available: https://cra.org/ccc/wp-content/uploads/sites/2/2017/06/SmartandPervasiveHealth-White-Paper-June-2017.pdf

[17] Graham S, Estrin D, Horvitz E, Kohane I, Mynatt E, Sim I. Information Technology Research Challenges for Healthcare: From Discovery to Delivery. Workshop Report on Discovery and Innovation in Health IT Workshop sponsored by the National Science Foundation, the Office of the National Coordinator for Health Information Technology, the National Institute of Standards and Technology, the National Library of Medicine, the Agency for Healthcare Research and Quality, the Computing Community Consortium, and the American Medical Informatics Association [Internet]. CCC; 2009. Available: http://archive2.cra.org/ccc/files/docs/init/Information_Technology_Research_Challenges_for_Healthcare.pdf

[18] Veinot TC, Ancker JS, Lyles C, Parker AG, Siek KA. Good Intentions Are Not Enough: Health Informatics Interventions That Worsen Inequality [Internet]. Workshop on Interactive Systems in Health Care (WISH) 2017; 2017 Nov 4; Washington, DC. Available: http://wish.gatech.edu/wp-content/uploads/ampanel.pdf

[19] Lorenc T, Petticrew M, Welch V, Tugwell P. What types of interventions generate inequalities? Evidence from systematic reviews. J Epidemiol Community Health. 2013;67: 190–193.

[20] Bränström R, Hatzenbuehler ML, Pachankis JE, Link BG. Sexual Orientation Disparities in Preventable Disease: A Fundamental Cause Perspective. Am J Public Health. 2016;106: 1109–1115.

[21] Hatzenbuehler ML, Phelan JC, Link BG. Stigma as a fundamental cause of population health inequalities. Am J Public Health. 2013;103: 813–821.

[22] Williams DR, Collins C. Racial residential segregation: a fundamental cause of racial disparities in health. Public Health Rep. 2001;116: 404–416.

[23] Phelan JC, Link BG. Is Racism a Fundamental Cause of Inequalities in Health? Annu Rev Sociol. 2015;41: 311–330.

[24] Solar O, Irwin A. A conceptual framework for action on the social determinants of health. Social Determinants of Health Discussion Paper 2 (Policy and Practice) [Internet]. World Health Organization; 2017 Aug. Report No.: Paper 2. Available: http://www.who.int/social_determinants/publications/9789241500852/en/

[25] Gibbons MC, Wilson RF, Samal L, Lehmann CU, Dickersin K, Lehmann HP, et al. Consumer health informatics: results of a systematic evidence review and evidence based recommendations. Transl Behav Med. 2011;1: 72–82.

[26] Thomas SB, Quinn SC. The Tuskegee Syphilis Study, 1932 to 1972: implications for HIV education and AIDS risk education programs in the black community. Am J Public Health. 1991;81: 1498–1505.

[27] Armstrong K, Putt M, Halbert CH, Grande D, Schwartz JS, Liao K, et al. Prior experiences of racial discrimination and racial differences in health care system distrust. Med Care. 2013;51: 144–150.

[28] Jimison H, Gorman P, Woods S, Nygren P, Walker M, Norris S, et al. Barriers and drivers of health information technology use for the elderly, chronically ill, and underserved. Evid Rep Technol Assess . 2008; 1–1422.

[29] Kim B, Han I. The role of trust belief and its antecedents in a community-driven knowledge environment. J Am Soc Inf Sci Technol. 2009;60: 1012–1026.

[30] Veinot TC, Campbell TR, Kruger DJ, Grodzinski A. A question of trust: user-centered design requirements for an informatics intervention to promote the sexual health of African-American youth. J Am Med Inform Assoc. 2013;20: 758–765.

[31] Unertl KM, Schaefbauer CL, Campbell TR, Senteio C, Siek KA, Bakken S, et al. Integrating community-based participatory research and informatics approaches to improve the engagement and health of underserved populations. J Am Med Inform Assoc. 2016;23: 60–73.

[32] Short CE, Vandelanotte C, Dixon MW, Rosenkranz R, Caperchione C, Hooker C, et al. Examining participant engagement in an information technology-based physical activity and nutrition intervention for men: the manup randomized controlled trial. JMIR Res Protoc. 2014;3: e2.

[33] Alfonsson S, Olsson E, Hursti T. Motivation and Treatment Credibility Predicts Dropout, Treatment Adherence, and Clinical Outcomes in an Internet-Based Cognitive Behavioral Relaxation Program: A Randomized Controlled Trial. J Med Internet Res. 2016;18: e52.

[34] Strecher V, McClure J, Alexander G, Chakraborty B, Nair V, Konkel J, et al. The Role of Engagement in a Tailored Web-Based Smoking Cessation Program: Randomized Controlled Trial. J Med Internet Res. 2008;10: e36.

[35] Nash CM, Vickerman KA, Kellogg ES, Zbikowski SM. Utilization of a Web-based vs integrated phone/Web cessation program among 140,000 tobacco users: an evaluation across 10 free state quitlines. J Med Internet Res. 2015;17: e36.

[36] Meischke H, Lozano P, Zhou C, Garrison MM, Christakis D. Engagement in "My Child's Asthma", an interactive web-based pediatric asthma management intervention. Int J Med Inform. 2011;80: 765–774.

[37] Murray E, White IR, Varagunam M, Godfrey C, Khadjesari Z, McCambridge J. Attrition Revisited: Adherence and Retention in a Web-Based Alcohol Trial. J Med Internet Res. 2013;15: e162.

[38] Robroek SJW, Lindeboom DEM, Burdorf A. Initial and Sustained Participation in an Internet-delivered Long-term Worksite Health Promotion Program on Physical Activity and Nutrition. J Med Internet Res. 2012;14: e43.

[39] Van 't Riet J, Crutzen R, De Vries H. Investigating Predictors of Visiting, Using, and Revisiting an Online Health-Communication Program: A Longitudinal Study. J Med Internet Res. 2010;12: e37.

[40] Kure-Biegel N, Schnohr CW, Hindhede AL, Diderichsen F. Risk factors for not completing health interventions and the potential impact on health inequalities between educational groups - a mixed method study from Denmark. Int J Equity Health. 2016;15: 54.

[41] Cortés YI, Arcia A, Kearney J, Luchsinger J, Lucero RJ. Urban-Dwelling Community Members' Views on Biomedical Research Engagement. Qual Health Res. 2017;27: 130–137.

[42] MacLeod H, Jelen B, Prabhakar A, Oehlberg L, Siek K, Connelly K. A Guide to Using Asynchronous Remote Communities (ARC) for Researching Distributed Populations. EAI Endorsed Transactions on Pervasive Health and Technology. 2017;3: 152898.

[43] Hayes GR. The relationship of action research to human-computer interaction. ACM Trans Comput-Hum Interact. 2011;18: 1–20.

[44] Simonsen J, Robertson T. Routledge International Handbook of Participatory Design. Routledge; 2012.

[45] Saxena A. Workforce Diversity: A Key to Improve Productivity. Procedia Economics and Finance. 2014;11: 76–85.

[46] Klasnja P, Consolvo S, Pratt W. How to evaluate technologies for health behavior change in HCI research. Proceedings of the 2011 annual conference on Human factors in computing systems - CHI '11. New York, New York, USA: ACM Press; 2011. p. 3063.

[47] Bowen DJ, Kreuter M, Spring B, Cofta-Woerpel L, Linnan L, Weiner D, et al. How we design feasibility studies. Am J Prev Med. 2009;36: 452–457.

[48] Klasnja P, Hekler EB, Korinek EV, Harlow J, Mishra SR. Toward Usable Evidence: Optimizing Knowledge Accumulation in HCI Research on Health Behavior Change. Proceedings of the 2017 CHI Conference on Human Factors in Computing Systems - CHI '17. New York, New York, USA: ACM Press; 2017. pp. 3071–3082.

[49] Mohr DC, Lyon AR, Lattie EG, Reddy M, Schueller SM. Accelerating Digital Mental Health Research From Early Design and Creation to Successful Implementation and Sustainment. J Med Internet Res. 2017;19: e153.

[50] Hekler EB, Klasnja P, Riley WT, Buman MP, Huberty J, Rivera DE, et al. Agile science: creating useful products for behavior change in the real world. Transl Behav Med. 2016;6: 317–328.

[51] Veinot TC, Mitchell H, Ancker JS. Good intentions are not enough: how informatics interventions can worsen inequality. J Am Med Inform Assoc. 2018; doi:10.1093/jamia/ocy052

[52] Hardy J, Veinot TC, Yan X, Berrocal VJ, Clarke P, Goodspeed R, et al. User acceptance of location-tracking technologies

in health research: Implications for study design and data quality. J Biomed Inform. 2018;79: 7–19.

[53] Davis R, Campbell R, Hildon Z, Hobbs L, Michie S. Theories of behaviour and behaviour change across the social and behavioural sciences: a scoping review. Health Psychol Rev. Taylor & Francis; 2015;9: 323.

[54] Gallo LC, Bogart LM, Vranceanu A-M, Matthews KA. Socioeconomic status, resources, psychological experiences, and emotional responses: a test of the reserve capacity model. J Pers Soc Psychol. 2005;88: 386–399.

[55] Wishik SM, Vynckt S. The use of nutritional "positive deviants" to identify approaches for modification of dietary practices. Am J Public Health. 1976;66: 38–42.

[56] Spruijt-Metz D, Nilsen W. Dynamic Models of Behavior for Just-in-Time Adaptive Interventions. IEEE Pervasive Comput. 2014;13: 13–17.

[57] Intille SS, Kukla C, Farzanfar R, Bakr W. Just-in-time technology to encourage incremental, dietary behavior change. AMIA Annu Symp Proc. 2003; 874.

[58] Schmid KL, Rivers SE, Latimer AE, Salovey P. Targeting or Tailoring? Maximizing Resources to Create Effective Health Communications. Mark Health Serv. NIH Public Access; 2008;28: 32.

[59] Clare Garvie JF. Facial-Recognition Software Might Have a Racial Bias Problem. In: The Atlantic [Internet]. 7 Apr 2016 [cited 4 Jul 2018]. Available: https://www.theatlantic.com/technology/archive/2016/04/the-underlying-bias-of-facial-recognition-systems/476991/

[60] Campbell M. Framework for design and evaluation of complex interventions to improve health. BMJ. 2000;321: 694–696.

[61] Collins LM, Murphy SA, Strecher V. The multiphase optimization strategy (MOST) and the sequential multiple assignment randomized trial (SMART): new methods for more potent eHealth interventions. Am J Prev Med. 2007;32: S112–8.

[62] Eikey EV. Unintended Users, Uses, and Consequences of Mobile Weight Loss Apps: Using Eating Disorders as a Case Study. In: Sezgin E, Yildirim S, Özkan-Yildirim S, Sumuer E, editors. Current and Emerging mHealth Technologies: Adoption, Implementation, and Use. Cham: Springer International Publishing; 2018. pp. 119–133.

[63] National Academies of Sciences, Engineering, and Medicine. Returning Individual Research Results to Participants: Guidance for a New Research Paradigm : Health and Medicine Division. Washington, DC: The National Academies Press; 2018.

[64] App Evaluation Model [Internet]. [cited 3 Aug 2018]. Available: https://www.psychiatry.org/psychiatrists/practice/mental-health-apps/app-evaluation-model

[65] Schick RS, Kelsey TW, Marston J, Samson K, Humphris GW. MapMySmoke: feasibility of a new quit cigarette smoking mobile phone application using integrated geo-positioning technology, and motivational messaging within a primary care setting. Pilot and Feasibility Studies. BioMed Central; 2017;4: 19.

[66] Naughton F, Hopewell S, Lathia N, Schalbroeck R, Brown C, Mascolo C, et al. A Context-Sensing Mobile Phone App (Q Sense) for Smoking Cessation: A Mixed-Methods Study. JMIR Mhealth Uhealth. 2016;4: e106.

[67] Jaimes L, Llofriu M, Raij A. A Stress-Free Life: Just-in-Time Interventions for Stress via Real-Time Forecasting and Intervention Adaptation. Proceedings of the 9th International Conference on Body Area Networks. 2014. doi:10.4108/icst.bodynets.2014.258237

[68] Saeb S, Zhang M, Karr CJ, Schueller SM, Corden ME, Kording KP, et al. Mobile Phone Sensor Correlates of Depressive Symptom Severity in Daily-Life Behavior: An Exploratory Study. J Med Internet Res. 2015;17: e175.

[69] Hung GC-L, Yang P-C, Chang C-C, Chiang J-H, Chen Y-Y. Predicting Negative Emotions Based on Mobile Phone Usage Patterns: An Exploratory Study. JMIR Res Protoc. 2016;5: e160.

[70] Gold R, Cottrell E, Bunce A, Middendorf M, Hollombe C, Cowburn S, et al. Developing Electronic Health Record (EHR) Strategies Related to Health Center Patients' Social Determinants of Health. J Am Board Fam Med. 2017;30: 428–447.

[71] Cantor MN, Thorpe L. Integrating Data On Social Determinants Of Health Into Electronic Health Records. Health Aff . 2018;37: 585–590.

[72] Dalton JE, Perzynski AT, Zidar DA, Rothberg MB, Coulton CJ, Milinovich AT, et al. Accuracy of Cardiovascular Risk Prediction Varies by Neighborhood Socioeconomic Position: A Retrospective Cohort Study. Ann Intern Med. 2017;167: 456–464.

[73] Sills MR, Hall M, Colvin JD, Macy ML, Cutler GJ, Bettenhausen JL, et al. Association of Social Determinants With Children's Hospitals' Preventable Readmissions Performance. JAMA Pediatr. 2016;170: 350–358.

[74] Fiesler C, Proferes N. "Participant" Perceptions of Twitter Research Ethics. Social Media+ Society. journals.sagepub.com; 2018;4: 1–14.

[75] Lai C. Engaging patient in health and health care processes: the role of patient platforms [Internet]. tspace.library.utoronto.ca. 2018. Available: https://tspace.library.utoronto.ca/handle/1807/92035

[76] Pérez-Stable EJ, Collins FS. Science Visioning in Minority Health and Health Disparities. Am J Public Health. 2019;109: S5.

[77] Pain E. A New Funding Model for Scientists. Science. 14 Jan 2015. doi:10.1126/science.caredit.a1400012

[78] Moher D, Naudet F, Cristea IA, Miedema F, Ioannidis JPA, Goodman SN. Assessing scientists for hiring, promotion, and tenure. PLoS Biol. Public Library of Science; 2018;16: e2004089.

## Workshop Attendees:

| First Name | Last Name | Affiliation |
|---|---|---|
| Sirry | Alang | Lehigh University |
| Suzanne | Bakken | Columbia University |
| Andrea | Barbarin | IBM Watson Health |
| Laura | Bartlett | NIH / National Library of Medicine |
| Elisabeth | Becker | University of Texas Health Science Center |
| Deanna | Befus | Wake Forest University |
| Marissa | Burgermaster | Columbia University |
| Cynthia | Castro-Sweet | Omada Health, Inc. |
| Yunan | Chen | University of California - Irvine |
| Heather | Cole-Lewis | Johnson and Johnson |
| Kay | Connelly | Indiana University |
| Mary | Czerwinski | Microsoft Research |
| Tawanna | Dillahunt | University of Michigan |
| Khari | Douglas | Computing Community Consortium |
| Valerie | Earnshaw | University of Delaware |
| Rachel | Gold | Kaiser Permanente Center for Health Research |
| Paul | Gorman | Oregon Health Sciences University |
| Syed | Haider | Johnson & Johnson |
| Eric | Hekler | University of California, San Diego |
| Sarah | Iribarren | University of Washington School of Nursing |
| Maia | Jacobs | Harvard University |
| Holly | Jimison | Northeastern University |
| Charles | Jonassaint | University of Pittsburgh Medical Center |
| Julie | Kientz | University of Washington |
| Katherine | Kim | University of California, Davis |
| Pedja | Klasnja | Kaiser Permanente Washington Health |
| Young-Ji | Lee | University of Pittsburgh |
| Robert | Lucero | University of Florida – Gainesville |
| Haley | MacLeod | Facebook |
| Lena | Mamykina | Columbia University |
| Gabriela | Marcu | Drexel University |
| Jessica | McCurley | Massachusetts General Hospital / Harvard Medical School |
| Sarah | Miller | Icahn School of Medicine At Mount Sinai |
| Enid | Montague | DePaul University |
| Brian | Mosley | Computing Research Association |
| Sarah | Mullane | Arizona State University |
| Sean | Munson | University of Washington |
| Elizabeth | Mynatt | Georgia Institute of Technology / CCC |

| Lyndsay | Nelson | Vanderbilt University Medical Center |
| Robert | Newton | Pennington Biomedical Research Center |
| Michelle | Odlum | Columbia University School of Nursing |
| Jessica | Pater | Georgia Institute of Technology |
| Misha | Pavel | Northeastern University |
| Wanda | Pratt | University of Washington |
| Madhu | Reddy | Northwestern University |
| Ashutosh | Sabharwal | Rice University |
| Charles | Senteio | Rutgers University School of Communication |
| Katie | Siek | Indiana University |
| Shawna | Sisler | University of Utah / UC San Francisco |
| Jamilia | Sly | Icahn School of Medicine at Mount Sinai |
| Stephanie | Sohl | Wake Forest School of Medicine |
| Bonnie | Spring | Northwestern University |
| Donna | Spruijt-Metz | University of Southern California |
| Michael | Stanton | California State University East Bay |
| Jasmin | Tiro | University of Texas, Southwestern Medical Center |
| John | Torous | Beth Israel Deaconess Medical Center / Harvard Medical School |
| Tammy | Toscos | Parkview Health |
| Kim | Unertl | Vanderbilt University Medical Center |
| Rupa | Valdez | University of Virginia |
| Tiffany | Veinot | University of Michigan |
| Lauren | Wilcox | Georgia Institute of Technology |
| Xinzhi | Zhang | NIH, NIMHD |

**NOTES**

# NOTES

**NOTES**

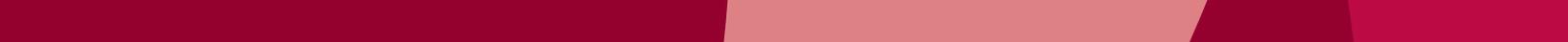

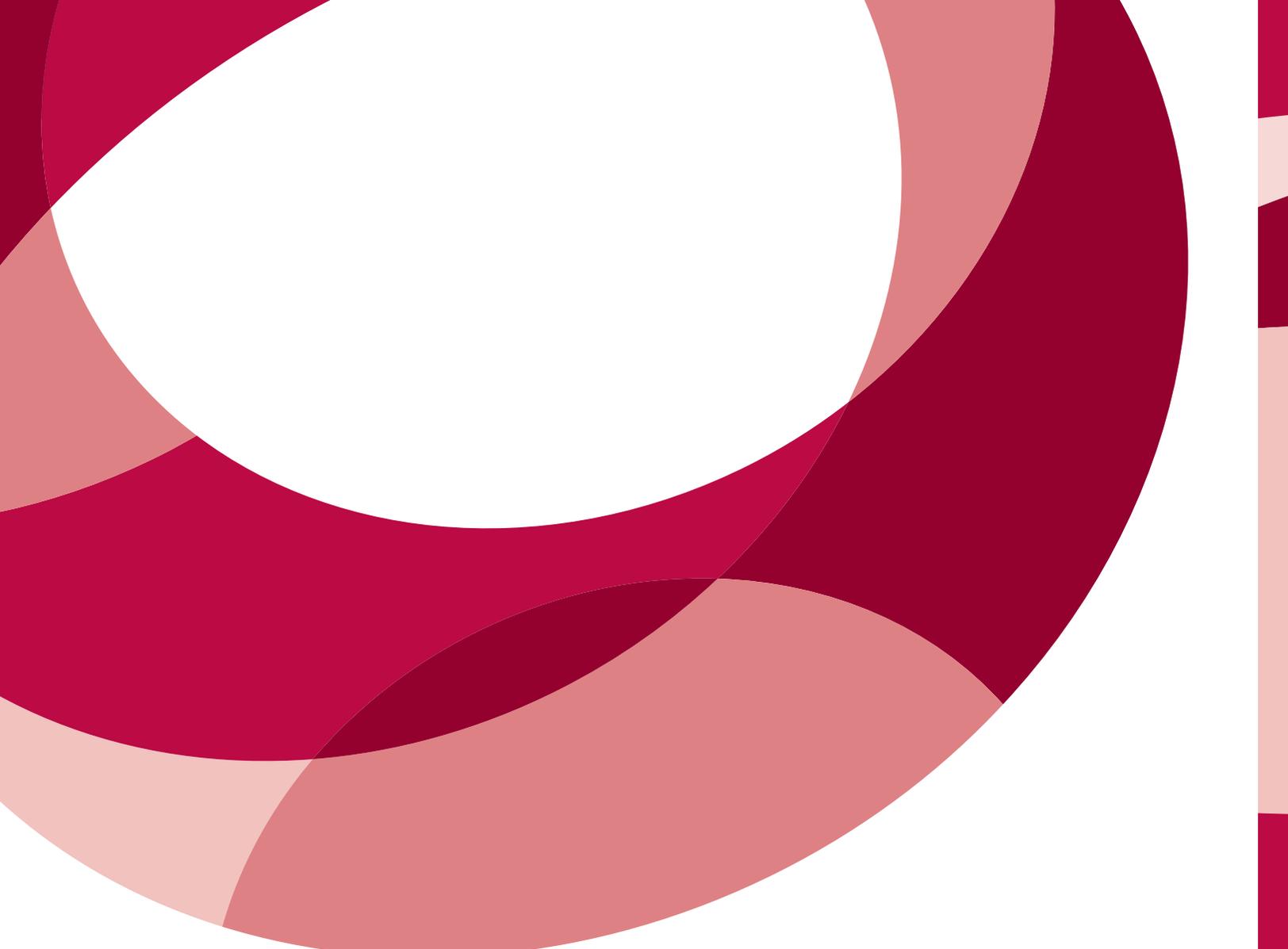
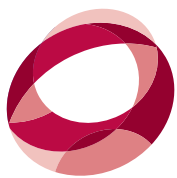